# Is Science Nearing Its Limits ? Summarizing Dialogue

*Jean-Pierre Luminet*
*Laboratoire Univers et Théories, CNRS-UMR 8102,*
*Observatoire de Paris, F-92195 Meudon cedex, France*
jean-pierre.luminet@obspm.fr

**Abstract**

In 2007 an international conference engaged a reflection on the present conditions for sciences and scientific practice, to initiate a dialogue between science, philosophy, epistemology and sociology. Starting from many examples taken from the history of sciences and the present status of astrophysics, cosmology and fundamental physics, I analyze various issues such as the possible limits of theoretical and technological advances, the paradox of non-verifiability in string theory, incompleteness, the role of imagination in scientific research along with educational aspects, the epistemological value of the Ockham's razor principle for constraining theories, and our undersanding of man's place in Creation.

**Is the Progress of Science Limited ?**

First of all I am rather sceptical about the idea suggested by George Steiner in his introductory lecture, namely that 'the progress of the sciences is limitless' works as a postulate of western rationality. At several key moments of the history of science – generally preceding or following shortly a decisive change of paradigm – it has been the opposite idea that prevailed, namely that the progress of science had reached its limits. I'll recall three historical examples.

The first one is linked to the extraordinary step in our capacity to understand the physical world following the Newtonian revolution in physics and astronomy. The progress was so spectacular that, in the following decades, it seemed that nothing better could ever be achieved in the field of science, implying that science was in some sense near its end. For instance, at the beginning of nineteenth century, the French mathematician Lagrange declared enthusiastically: 'Since there is only one universe to explain, nobody can redo what Newton did, happiest of the mortals'.[1] Another characteristic quotation is from the poet Gudin de la Brunellerie (a close collaborator with the famous Beaumarchais) : 'Secrets of the Almighty until Newton veiled / It is Newton who revealed you to the mortals.' This is my bad English translation of a bad French poem![2]

In the same epoch, the so-called 'French Newton', namely Pierre Simon, marquis de Laplace, assumed that, due to the perfection of physics, the world was fully deterministic ; as he wrote in his *Essai philosophique sur les probabilités*: 'We may regard the present state of the universe as the effect of its past and the cause of its future. An intellect which at a certain moment would know all forces that set nature in motion, and all positions of all items of which nature is composed, if this intellect were also vast enough to submit these data to analysis, it would embrace in a single formula the movements of the greatest bodies of the universe and those of the tiniest atom; for such an intellect nothing would be uncertain and the future just like the past would be present before its eyes.'[3] So, if the world is fully deterministic, what is left for the imaginative scientist? Note that Laplace's conception prevailed a long time, even when new theoretical breakthroughs such as quantum mechanics or chaos theory showed the limits of determinism. This wrong conception is probably responsible for the deep misunderstanding of what science really is, leading for instance to the false idea that 'science is only aimed at providing answers and not at positing questions' (Heidegger, as quoted by George Steiner in his introductory lecture).

The second example is commonly quoted: at the end of the nineteenth century, following decisive new progress in physics due to Maxwell's theory of electromagnetism, the British physicist William Thomson, Lord Kelvin, gave a lecture entitled 'Nineteenth-Century Clouds over the Dynamical Theory of Heat and Light'.[4] The two 'dark clouds' he was alluding to were the unsatisfactory explanations that the physics of the time could give for two phenomena: the Michelson-Morley experiment about the absolute motion of the Earth in the ether, and the unexplained properties of black body radiation. But Kelvin predicted that these two dark clouds would be quickly dissipated in the framework of classical physics, thus signalling an end of physics. As it is well-known, it happened that, starting from these clouds, two revolutionary physical theories were developed during the twentieth century: the theory of relativity and quantum mechanics.

The third example is directly linked to the present status of theoretical physics. In the 1970s, many particle physicists were engaged in developing a theory that purported to unify all the fundamental forces of nature. Such a theory was overzealously dubbed the Theory of Everything (TOE). A number of professionals were embroiled in hot discussions, suggesting that such a theory might herald an end to physics. The influential and popular Stephen Hawking was so enthusiastic about the so-called N=8 supergravity theory that, opening his Lucasian lecture in 1980, he declared 'I want to discuss the possibility that the goal of theoretical physics might be achieved in the not too distant future, say by the end of the century.'[5] Such a statement implied that there might not be any challenging and exciting problems left for the physicists to work on after TOE. As you know, there is no TOE yet available, and Stephen Hawking had recently to change his mind about the end of physics, by taking into account Gödel's incompleteness theorem – although, in my opinion, again in an unsound manner, as I shall comment later.

Thus, the topic of the present conference, 'Is Science Nearing Its Limits?', is far from being a new question. However, the present questioning has its own originality compared to previous ones, because it rests on a quite different line of reasoning. The previous questionings assumed that science was near its limits because of its apparent triumphal success. When extraordinary achievements are made in a given discipline, one can have the impression that it will be hard to do better, and thus that we have reached the limits of human capacities (it is also the case in sporting competition, for instance, world records in athletics). The present questioning assumes, on the contrary, that science could be near its limits because of its dismal failure. Is this complete change of view just a sign of the general pessimism that can be felt in practically all the fields of human activity in the western world during the first decade of our century, or does it rest on a real crisis of science?

We heard about different opinions during this workshop, and the issue is certainly not clear. My own opinion, for what it's worth, is that science, far from being near its end, is *constantly near to its beginnings!* Of course there are fields of science more developed than others, because they are in some way less complex, less entangled with many unmanageable parameters. This is the case for the so-called exact sciences. But I feel that neural science or cognitive sciences, for instance, are only in the infancy of understanding theoretically the systems they study, namely the human brain and conscience. Yesterday morning I should have said that their present theoretical status is comparable to the status of astronomy and physics at the time of Galileo and Kepler, and that the major revolutions are to come. Now I'm not so sure after hearing the brilliant lectures by Gerald Edelman and Wolf Singer yesterday afternoon: perhaps the right correspondance would be with the time of Newton.

**Are there Technological Limits ?**

Now I would like to comment shortly on some directions of debate put forward by George Steiner in his opening conference.

I do not think that our means of observation are nearing their technological limits. For instance, whatever the progress of electronic microscopy, it is out of question to reach in a foreseable future a spatial resolution of the order of the Planck length scale, $10^{-33}$ cm, both for fundamental reasons, linked to quantum fluctuations of space-time itself, and due to technological limitations. The latter come from the fact that in particle physics, the distance scales are related to energy scales: the more one gives a particle a great energy to collide with other particles at speeds close to that of light, the more finely one probes the space extension of this particle, and the more one tests the subtle structure of space. Therefore the subatomic scale of space is now better explored by particle accelerators than by electronic microscopes. But even with this enhancement, the length scales now reached by the most powerful accelerators is only $10^{-18}$ cm, extraordinarily far for the Planck length limit. In order to reach the Planck length we would need particle accelerators of the size of the universe. Clearly it is not for tomorrow! The fact we are so far from the limits is a pity, because if we could test the Planck length, we could test experimentally quantum gravity theories, such as superstring theory. But as you know, the non-demonstrability of superstring theory due to such an impossibility has been one of the topics of this conference.

The situation is similar at the opposite end of the scale, namely in cosmology and astrophysics. It is true that some of our astronomical tools, such as radiotelescopes and optical instruments, are nearing the confines of observable space; still better, the COBE and WMAP microwave telescopes have already reached the farthest possible source of electromagnetic radiation, namely the cosmic black body radiation. It was emitted 14 billions years ago by the primordial plasma, more precisely 400,000 years after the Big Bang, and the corresponding light source is now situated at a distance of about 50 billion light-years all around us – the present-day distance of the cosmological horizon. This looks an impressive panorama. However, such a panorama embraces only the electromagnetically observable universe. Yet there are sources of information other than electromagnetic radiation, for instance neutrinos and gravitational waves, which can potentially provide us with a much wider exploration of space-time at a large scale, and our technology in these fields is still in its infancy. We have already some detectors able to catch high-energy neutrinos emitted by the centre of the Sun or by supernovae explosions, but we are very far from having detectors able to grasp the much lower-energy neutrinos emitted by the primordial universe. Again, it is a pity, because if we could detect the cosmic neutrino background, we could directly probe the universe when it was only one second old (after the Big Bang). It is also a source of great excitement for the future, since it involves several decades of technological improvements in order to reach the limit, and frankly I don't see what could limit such technological improvements.

The universe is also filled with various sources of gravitational waves, such as black hole vibrations, collisions between compact stars, supernovae explosions, and the Big Bang. According to relativistic cosmology, primordial gravitational waves were emitted when the universe was only $10^{-43}$ second old. Still more exciting, in some quantum gravity models, the Big Bang itself as a space-time singularity is smoothed out and does not mark the beginning of time; in other words, you can now ask the question 'What happened before the Big Bang?' without risking an arrogant reply from a professional cosmologist pretending that your question is a nonsense; you can also avoid Augustine's more humourous reply to those who asked the similar question, 'What was God doing before the creation of the universe?' Well, God was preparing Hell for those who dared ask!

Now, several quantum gravity theories predict a pre-Big Bang era of the universe, and specific gravitational waves would be the observational imprint of such an era. Their detection could help to distinguish between several theoretical proposals, for instance those coming from superstring theories and those coming from quantum loop gravity (an alternative road to quantum gravity).

However the experimental status of gravitational waves is still at an extremely preliminary stage; the presently available detectors, like the VIRGO and LIGO interferometers, are potentially able to detect only very strong signals emitted by black holes or neutron star coalescences in the local supercluster of galaxies; yet no such signal was detected after two years of activity. The next generation interferometer, namely the LISA project put in orbit around the Sun, will improve the sensitivity, but will remain extremely far from the possibility of detecting low-amplitude gravitational waves of cosmological origin. Again, this is both a source of frustration for theorists, and a source of great excitement for experimentalists, who have decades of future work in order to improve their instruments.

**The Karl Popper's Criteria**

Perhaps I have been a little bit too optimistic about the improvements of our experimental tools, so let us assume for a moment that our technology will not progress much more in the next decades, and assume that we will not be able to test experimentally some theoretical expectations at high-energy or very small scales. Clearly such theoretical expectations would no more obey the usual Popperian criteria of predictability and falsifiability;[6] is it really a crisis of science?

Again the history of science shows us how, at several moments, the theories could already be so far in advance of the available instrumental tools that at the time they were proposed, it was virtually impossible to hope for an experimental check, just because the corresponding technology was not even thinkable. Two classic examples are easily found in Greek science. One concerns the smallest scales, with the atomistic theory of Democritus and Epicurus, the other concerns the astronomical scale with the heliocentric theory of Aristarchus of Samos. Was it science or not? According to Popper's criteria, it was not, because at the time these theories were put about, the scale of atoms was totally out of experimental reach, as well as the astronomical proofs about the motion of the Earth. However, many centuries later, the atomistic and heliocentric theories returned to the front of science and became the sources of two major scientific revolutions. Regarding the atomistic theory, the first experimental indications had to wait until the nineteenth century and John Dalton. Regarding the heliocentric theory, the first experimental proof for the motion of the Earth came from James Bradley in 1727, namely, still well after the resurrection of the Aristarchus theory by Copernicus and Galileo. Thus even Copernicus and Galileo did not produce science? These well-known examples suggest that the Popperian criteria of what is science and what is not might be revised and enlarged.

Take now the widely discussed status of unification theories, string theory or quantum loop gravity. They make predictions at scales not reachable by presently available experimental tools.

Sure, they will never be tested in particle accelerators. But it might just be that nobody was imaginative enough to foresee the corresponding technology… In that sense, I am rather confident that quantum loop gravity is a promising scientific theory. I shall be less confident with string theory, because the crucial point concerning its non-demonstrability does not lie in the fact that it deals with scales inaccessible in a foreseeable future to experimental checks, but in the fact that there is not *one* string solution, as brightly explained yesterday by Dieter Lüst and Peter Voit, there are about $10^{500}$ different string solutions, each one making different predictions about the world picture. This looks really crazy, since even if we had the technology to test the energy scales of strings, it would be impossible to distinguish between so many models, all working at the same scale. You have heard how some fans of string theory escape the problem by promoting the idea of a 'landscape';[7] for them, the theoretical stringy framework is a kind of immanent truth; thus the various models have no more to be adapted to and constrained by the observed properties of the universe. On the contrary, our conception of the physical universe should be adapted to the conceptual framework; as a consequence, each one of the $10^{500}$ string solutions would be true, and there should exist a corresponding specific universe within a so-called multiverse!

This is certainly a very provocative and disputable proposal, I myself feel very cautious about such a landscape theory, but I must recognize that the proposal has the merit to put into question the criteria put forward by the 'Popperazzi'. After all, the Popper criterion of falsifiability is itself a theory (in the field of epistemology and philosophy of knowledge) and, as goes for any other theory, I do not see why it should be definitively accepted and undisputed.

**Incompleteness**

Another barrier to scientific knowledge put forward by various speakers of the conference refers to Gödel's incompleteness theorems (I put an 's' because there are two of them).[8] Some thinkers have argued that the theorems have implications in wider areas of philosophy and even cognitive science. For instance, John Lucas[9] and Roger Penrose[10] used them to suggest that we could never reduce brain activity to computational algorithms and the human mind to a Turing machine. They argued that if it was so, and if the machine was consistent, then Gödel's incompleteness theorems would apply to it. On the same basis, Stanley Jaki,[11] followed much later by Stephen Hawking,[12] argued that even the most sophisticated formulation of physics will be incomplete, and that therefore there can never be an ultimate theory that can be formulated as a finite number of principles.

Appeals were sometimes made to the incompleteness theorems to support by analogy ideas which go well beyond mathematics and logic. For instance, the French intellectual Régis Debray applied it to politics![13]

Happily, a number of authors have commented negatively on such extensions and interpretations. Although Gödel's theorems are an extraordinary result of mathematics, I am not sure that their real domain of application is as large as so many thinkers believe. Gödel's theorems just prove that Hilbert's programme to find a complete and consistent set of axioms for all of mathematics is impossible. For instance, Gödel's theorems cannot be applied to humans, since the latter make mistakes and are therefore inconsistent.

After all, Gödel's theorems concern formal systems, and if it is true that any axiomatic system contains at least one undecidable proposition, it is also true that it is possible to build up a new, larger logical system in which the previously non-demonstrable statement is assumed to be true, or false, and incorporated as an additional axiom. Of course, in turn the new logical system will have its own new undecidable proposition, and this is the most fascinating part of Gödel's theorems. But it also suggests that they do not necessarily impose an absolute limit to the laws of thought, only a moving limit. And as you know, a moving limit is not an end.

**What is Science ?**

I give now a short personal comment about Heidegger's *boutade*: 'The sciences are trivia; all they give us is answers. It is the questions which matter.' If I fully agree with the second part, I regret to say that the first part of the statement is a complete misunderstanding about what science really is. Any creative researcher asks for questions, and not for answers. Go to a classroom in a college, only a tiny fraction of the pupils will later become professional scientists, and it will be almost impossible to detect in advance who will become a professional scientist. But at least you can surely select a subset of the pupils, those who sometimes ask questions of their teacher, and who do not blindly accept the teaching as indisputable answers. The same holds for professional researchers. At least in my field, theoretical physics, our deepest motivation is to raise new questions rather than to provide answers. I am not sure that it is the same in experimental physics, because experiments are necessarily designed to provide answers to well-posed questions (otherwise it would be impossible to build the adequate apparatus!). Thus, it looks as though for Heidegger, along with too many other philosophers who do not have a deep scientific background, and still worse along with most of politicians in charge of education and research, science was reduced to its experimental part. This is a major mistake, and such a misunderstanding about the real aims of science might be one of the causes of the sharp decline in student enrolment in science. I myself have many times gone some way to counterbalancing the decline a little by explaining enthusiastically how science is not just a dead boring discipline where people spend their time checking reality with big instruments, but is also and essentially an extraordinarily creative field, a ludic activity calling for all the resources of the imagination, even sometimes of the dream. (Since I am not completely naive, I also think that a major cause of the decline of science in western institutions is linked to the low level of salaries, taking account of the considerable talent and efforts it requires to become a professional…)

**Educational Aspects.**

In a recent book,[14] the physicist Lee Smolin pointed out some dysfunctions of academic research and rightly analysed them within the framework of the sociology of science. One of these dysfunctions is why certain subjects can monopolize the resources available for research and, consequently, cut the resources available to the exploration of other approaches of comparable, if not higher promise. Such is the case of string theory, which has monopolized the time and the energy of so many researchers for at least thirty years.

The sociology of science naturally wants it to be, that at each moment of history, there is a field dominating inside theoretical physics. In the 1930s it was nuclear physics, then in the 1960s it was particle physics. String theory is the most recent example of a dominating field. However, it is distinguished from the preceding fields by its incapacity to make experimental predictions. I can note that in my own field of cosmology, the theory of inflation has played a similar role since the 1980s; it is almost impossible to be taken seriously as a professional cosmologist if you do not adhere blindly to inflation theory, and any new experimental result is interpreted as a confirmation of the theory, although the latter is not falsifiable, due to its large number of freely adjustable parameters. In fact, one could ironically say that the only thing proven by inflation is the following theorem: 'inflation can prove everything'.

In this respect, string theory or inflation give bad examples for those who would like, rightly, a reform of university programmes. University programmes are quite generally eighty years old, out-of-date compared to contemporary science. But if contemporary science is feeding ammunition to its enemies while putting on a pedestal theories which can neither be proven nor refuted, the adversaries of a reform of the programmes will find arguments to support their static point of view.

### The Role of Imagination

There are two kinds of researchers: some for normal science, others for revolutionary science. For the normal periods (say within a paradigm), one needs researchers who work effectively by controlling all the technical tools; they are 'master craftsmen'. Today, 95 per cent of the researchers in string theory are master craftsmen. It is they who were always the best students in maths and physics, from college until the PhD thesis, able to solve mathematical problems more quickly and better than the others.

But for the revolutionary periods, one needs visionaries. Einstein was one of them, like Niels Bohr. Kepler and Newton are very rare examples who had both qualities. The visionaries decide to do science because they raise questions which their textbooks do not answer. If they had not become scientists, they could have become painters, writers or musicians. And indeed there are many similarities between artistic and scientific creativity. I'll just recall the famous Einstein quotations: 'Imagination is more important than knowledge. Knowledge is limited. Imagination encircles the world'[15], and also 'Man tries to make for himself in the fashion that suits him best a simplified and intelligible picture of the world; he then tries to some extent to substitute this cosmos of his for the world of experience, and thus to overcome it. This is what the painter, the poet, the speculative philosopher, and the natural scientists do, each in his own fashion.'[16]

### The Principle of Simplicity

Now, I would like to develop a theme concerning one of the criteria often invoked to justify the domination of such or such theory over competing theories.

One of the peremptory arguments sometimes put forward by string theorists to impose their view as the only way worthy of research is that string theory would be more elegant, more beautiful and, in a certain way, simpler, since it aims to unify all the fundamental interactions into a single theory, the TOE (remember the title of Brian Greene's best-seller on string theory[17]).

Again, this is not new. In the fourteenth century, the English philosopher William of Ockham wrote: 'It is useless to achieve by a greater number of means what a less number of means is enough to produce [...] When things must make true a proposal, if two things are enough to produce this effect, it is superfluous to put three of them.'[18] In other words, Ockham says that, for a range of models that explain some given facts, preference should be given to that one which invokes the minimal number of hypotheses.

Throughout the history of thought, from Antiquity to the most recent developments of physics and cosmology, such a 'principle of simplicity', also called 'Ockham's razor', played a key role in the development of scientific, philosophical, even economic models (indeed, influenced by the Anglo-Saxon pragmatic thought, it is also called the 'principle of economy'). It even was at the source of true scientific revolutions.

Let us take the example of the theory about the motion of the Earth. Aristotle supposed a motionless Earth at the centre of the universe, all the celestial bodies, Sun included, being supposed to turn around it in twenty-four hours. The Greek astronomer, Aristarchus of Samos, claimed however that, the Sun being much larger than the Earth, it was simpler to suppose the small Earth as moving around the huge fixed Sun.

It is, however, the Aristotelian system which triumphed over many centuries. The hypothesis of a moving Earth reappeared only in the fourteenth century. Then, the philosopher Jean Buridan used the principle of simplicity to declare: 'It is better to account for appearances by little than by much of causes, if possible, and the same to give an account of it by the easiest way. However, it is easier to move what is small than what is large; it is thus better to say that the earth, which is small, is moved very quickly and that the supreme sphere [the Sun] is motionless, rather than saying the opposite.'[19]

In his famous treatise *De revolutionibus orbium coelestium* (1543), Nicolaus Copernicus argued in the same way: 'At rest, however, in the middle of everything is the sun. For in this most beautiful temple, who would place this lamp in another or better position than that from which it can light up the whole thing at the same time ?'[20]

Two other great revolutionaries of the history of sciences, Newton and Einstein, also made call on several occasions on the principle of simplicity to support their new visions of the world. More recently, it is thanks partly to the principle of simplicity that the models of 'black holes' and those of the 'Big Bang', now at the heart of astrophysics and modern cosmology, became credible: indeed, compared with competitor models, they better succeed in explaining the observations with the minimal number of assumptions.

The concept of simplicity in physics is also related to an aesthetic view: 'It seems that if one is working from the point of view of getting beauty into one's equation, and if one has really a sound insight, one is on a sure line of progress', the physicist Paul Dirac wrote.[21] 'It is a mysterious thing in fact how something which looks attractive may have a better chance of being true than something which looks ugly' emphasized Roger Penrose[22], one of the most imaginative contemporary scientists.

However, if the principle of simplicity is aesthetically appealing, its application is delicate insofar as the concept of simplicity relates to a state of knowledge at a given time. The history of sciences swarms with examples where arguments of simplicity, altogether subjective because they are very dependent on a framework of reasoning, impose a kind of preference, even of a fashion, and are able to achieve some form of orthodoxy, by excluding all models apparently less simple! The English scientist Francis Bacon (seventeenth century) had already remarkably formulated this objection: 'The human understanding is of its own nature prone to suppose the existence of more order and regularity in the world than it finds. And though there be many things in nature which are singular and unmatched, yet it devises for them parallels and conjugates and relatives which do not exist.'[23]

The immoderate use of the principle of simplicity in physics may thus generate perverse effects. The empirical success of a model creates a new aesthetic 'canon' of what must be 'simplicity'. This new canon is next used to justify theoretical fashions or diktats, which are not always experimentally testable. By creating a kind of 'soft' consensus in the scientific community, it is transformed into orthodoxy and ends up becoming . . . counter-revolutionary!

It is currently the case with string theory, which claims to be an elegant theory of unification of all four physical interactions, but which cannot be refuted nor proved by any experiment. Its alleged elegance is used, however, as an argument of authority, so that hundreds of young researchers in the world are forced to work on it, while other approaches are considered as marginal and depreciative.

Indeed, application of the principle of simplicity is delicate, since every as yet unknown phenomenon could be a source of 'natural' supplementary parameters. The notion of simplicity is therefore relative to the state of knowledge at a given moment. For instance, we have long believed that natural space was three-dimensional Euclidean space; general relativity taught us that space-time is naturally curved by gravity. Is Euclidean space simpler than non-Euclidean space ? At first view yes, but it does not fit the real structure of the universe.

Another example is taken from relativistic cosmology: should one prefer models of the universe without a cosmological constant to those that have one? Introduced by Einstein for a reason which was soon after revealed to be erroneous – namely to force relativistic models of the universe to remain static[24],– this cosmological constant subsequently seemed superfluous, to the point that solutions which went without it have drawn the almost exclusive attention of researchers for decades. However, following recent developments, both theoretical and observational, the situation has brusquely reversed. Today, the cosmological constant or an equivalent form known as 'dark energy' appears as a fundamental necessity of quantum theory and seems to have been confirmed by observational data. Necessity and opportunism make the law: if we need a cosmological constant to render our models of the universe coherent, is it not more 'simple' to assume it exists? This is an example where the strict application of the principle of simplicity has proven to be unfruitful, and has even hampered the development of the field.

In my own work as researcher, when I worked out a possible 'wraparound' shape of space,[25] I also encountered arguments of authority coming from the community: why, said most of my colleagues, imagine complex shapes and topologies of space (for example, a space which is finite, without edge and multiply connected, as I proposed), when it is simpler to suppose it flat and infinite?

It is true that the topological parameters (such as the structure of the fundamental polyhedron and the composition of the holonomy group) introduce extra factors in the standard cosmological models. At first sight, it may therefore seem wise to opt for the simply connected topology, all the more so since the corresponding models are easier to deal with. But to suppose because of this that the most easily handled theories have more of a chance of being true than others is to make a metaphysical hypothesis on the simplicity of nature, without having objectively specified what is meant by the degree of simplicity. If the topological parameters should be demanded by a physical theory which is deeper than general relativity, the argument of simplicity in favour of a simply connected topology would lose all value. In this kind of debate, the experiments must have the last word, and not subjective criteria of simplicity. In 2003, precise observations obtained by the NASA satellite WMAP revealed anomalies compared to the orthodox model of a flat and infinite space.[26] Thanks to these observations, we could publish an article which made the cover story of the journal *Nature*, with a precise proposal for a wraparound space: the dodecahedral spherical space.[27] Useless to specify that this model, mathematically complex but explaining remarkably well the observations, appears more elegant to me, and finally more 'simple', than the orthodox model. . .

**Man's Place in the Universe**

What we see in the skies poses the impression to our eyes that the skies were always similar, so that they did not have a beginning, as they will not have an end. However, the feeling of our absolute cosmic safety exists only because of our very ephemeral duration.

Beyond the myths and wild imaginings which man has always forged to build a comprehensible and reassuring picture of the universe, the modern cosmologist has observational facts and coherent theoretical interpretations, which enable him to reconstitute the past history of the universe and to calculate its future.

Cosmic prospects on the ends of our physical world may give the impression that the fate of mankind is so localized in time and space that it presents no interest. The inescapable end of mankind does not have any species of influence on the end of the cosmos. On the contrary, the inevitable flashover of our planet, the inevitable extinction of our star, the inevitable disintegration of organic matter, seal in the more or less short term the destiny of mankind.

It is undoubtedly what the philosopher Bertrand Russell, at the very beginning of twentieth century, wanted to express, at a time when the second law of thermodynamics triumphed, ensuring that any physical system was ineluctably dedicated to the degradation of energy, growth of entropy and disorganization: 'That all the labours of the ages, all the devotion, all the inspiration, all the noonday brightness of human genius, are destined to extinction in the vast death of the solar system, and that the whole temple of Man's achievement must inevitably be buried beneath the debris of a universe in ruins – all these things, if not quite beyond dispute, are yet so nearly certain that no philosophy which rejects them can hope to stand.'[28]

But is such a 'realistic' pessimism really appropriate? The eschatological scenarios, be they of a scientific, philosophical or religious nature, are all secretions of human thought, and are thus basically anthropomorphic. This is why some twentieth-century thinkers and futurologists, such as Freeman Dyson,[29] wanted to widen the debate, while wondering about a possible survival in the very long term of what seems to be the very essence of existence, namely conscience and intelligence.

Especially, the universe is not reduced to scientific investigation only. Now I conclude not only as a scientist but as a humanist. Art, poetry and, sometimes, philosophy probe the world to a depth that no telescope, no equation can reach. As Ernst Jünger wrote, 'There will always be men who hold the quality of time to be more important than its measurability. Nobody is unaware of this in the depth of his heart. Time does not provide only the framework of life; it is also the clothes of destiny. It does not mark only the limits to life, it is also its property. At the birth of each man emerges the time which is his.'[30]

There is a right refusal of humans to break the ropes which emotionally attach them to the cosmos. For the spontaneous glance, the Earth is flat, the Sun turns around and the starry sky is a vault which covers it. For these distorted, but concrete and simple realities, science substituted less distorted, but more abstract and complex realities. Science would be disavowed if it counterfeited its method by offering an image of the world in conformity with our desires. Does the feeling of plenitude that man could formerly attest under the starry vault have now to be destroyed like matter and energy in black holes, torn apart like expanding space by dark energy? To those who would be tempted to believe in such a dull conclusion, I will recall the masterly lesson of Blaise Pascal,[31] opposing the depth of thought to the extents of space and time:

> *Thought is the real measure of man's size. Man is but a reed, the most feeble thing in nature, but he is a thinking reed. The entire universe needs not arm itself to crush him. A vapour, a drop of water suffice to kill him. But, if the universe were to crush him, man would still be more noble than that which killed him, because he knows that he dies and the advantage which the universe has over him, the universe knows nothing of this. All* *our dignity then, consists in thought. By it we must elevate ourselves, and not by space and time which we cannot fill. Let us endeavour then, to think well; this is the principle of morality.*

---

[1] 'Comme il n'y a qu'un seul univers à expliquer, personne ne peut refaire ce qu'a fait Newton, le plus heureux des mortels.' Quoted in J.-P. Luminet, *Le Destin de l'Univers*, Paris: Fayard, 2006.
[2] ' Secrets du Tout-Puissant jusqu'à Newton voilés / C'est Newton aux mortels qui vous a révélés', from *L'astronomie* (Paris, 1810). Quoted in J.-P. Luminet, *Les poètes et l'univers*, Paris: Le Cherche midi, 1996.
[3] Pierre Simon de Laplace, *Essai philosophique sur les probabilités*, Paris, 1814. English translation by A. Dale, *Philosophical Essays on Probabilities*, New York: SpringerVerlag, 1995.
[4] *The London, Edinburgh and Dublin Philosophical Magazine and Journal of Science*, Series 6, volume 2 (1901), p.1.
[5] S. Hawking, *Black Holes and Baby Universes*, New York: Bantam Books, 1993, p. 49.
[6] K. Popper, *Logik der Forschung*, Vienna: Springer, 1934. Amplified English edition by K. Popper, *The Logic of Scientific Discovery*, London: Routledge, 1959.
[7] See, e.g., L. Susskind, *The Cosmic Landscape: String Theory and the Illusion of Intelligent Design*, London: Little, Brown, 2005.
[8] For an introduction, see E. Nagel and J.R. Newman, *Gödel's Proof*, London: Routledge, 1989 [1958].
[9] J.R. Lucas, *The Freedom of the Will*, Oxford: Clarendon Press, 1970.
[10] R. Penrose, *Shadows of the Mind*, Oxford: Oxford University Press, 1994.
[11] S. Jaki, *A Late Awakening to Gödel in Physics*, http://pirate.shu.edu/~jakistan/JakiGodel.pdf
[12] S. Hawking, 'Gödel and the End of Physics', talk at the Dirac Centennial Celebration, Cambridge, 20 July 2002. http://www.damtp.cam.ac.uk/strtst/dirac/hawking/
[13] See J. Bouveresse, *Prodiges et vertiges de l'analogie: de l'abus des belles-lettres dans la pensée*, Paris: Raisons d'agir, 1999, pp. 13-14.
[14] L. Smolin, *The Trouble With Physics: The Rise of String Theory, the Fall of a Science, and What Comes Next*. Boston, MA: Houghton Mifflin, 2006.
[15] *What Life Means to Einstein: An Interview by George Sylvester Viereck*, for the 26 October 1929 issue of the *Saturday Evening Post*.
[16] *Principles of Theoretical Physics*, address by Albert Einstein for Max Planck's sixtieth birthday, Physical Society, Berlin, 1918. Reprinted in A. Einstein, *Ideas and Opinions*, Pinebrook, NJ: Dell, 1954.
[17] B. Greene, *The Elegant Universe: Superstrings, Hidden Dimensions, and the Quest for the Ultimate Theory*, London: Vintage, 2000.
[18] The standard edition is Gedeon Gál, et al., (eds), *William of Ockham, Opera philosophica et theologica*, 17 vols, St Bonaventure, NY: The Franciscan Institute, 1967–88.
[19] J. Buridan, *Summulae de Dialectica*. English translation by G. Klima, New Haven/London : Yale University Press, 2001.
[20] *On the Revolutions of the Celestial Spheres*, translation and Commentary by E. Rosen, Blatimore, MD/London: The Johns Hopkins University Press, 1992.
[21] P. Dirac, 'The evolution of the physicist's picture of nature', *Scientific American* 208(5), 1963.
[22] R. Penrose, *The Role of Aesthetics in Pure and Applied Mathematical Research*, Bull. Inst. Maths. Appl. 10 (1974), p.266.